\documentclass[aps,prd,a4paper,twocolumn,amsmath,showpacs,superscriptaddress,nofootinbib,preprintnumbers]{revtex4-2}
\usepackage{graphicx}
\usepackage{epsf}
\usepackage{bm}
\usepackage{amsmath}
\usepackage{amsfonts}
\usepackage{amssymb}
\usepackage{epstopdf}
\usepackage{natbib}
\usepackage[dvipsnames,table]{xcolor}
\usepackage{physics}
\usepackage{hyperref}
\usepackage{lipsum}
\usepackage{float}
\usepackage{multirow}

\usepackage[capitalize]{cleveref}
\providecommand{\U}[1]{\protect\rule{.1in}{.1in}}

\newcommand{\be}{\begin{equation}}
\newcommand{\ee}{\end{equation}}
\newcommand{\bea}{\begin{eqnarray}}
\newcommand{\eea}{\end{eqnarray}}

\begin{document}

\title{Exploring the Hubble tension with a late time Modified Gravity scenario}

\author{Luis A. Escamilla}
\email{l.a.escamilla@sheffield.ac.uk}
\affiliation{School of Mathematics and Statistics, University of Sheffield, Hounsfield Road, Sheffield S3 7RH, United Kingdom}

\author{Donatella Fiorucci}
\affiliation{ENEA, Nuclear Department, C.R. Frascati, Via E. Fermi 45, 00044 Frascati (Roma), Italy}

\author{Giovanni Montani}
\affiliation{ENEA, Nuclear Department, C.R. Frascati, Via E. Fermi 45, 00044 Frascati (Roma), Italy}
\affiliation{Physics Department, “Sapienza” University of Rome, P.le Aldo Moro 5, 00185 Roma, Italy}

\author{Eleonora Di Valentino}
\email{e.divalentino@sheffield.ac.uk}
\affiliation{School of Mathematics and Statistics, University of Sheffield, Hounsfield Road, Sheffield S3 7RH, United Kingdom}

\begin{abstract}
We investigate a modified cosmological model aimed at addressing the Hubble tension, considering revised dynamics in the late Universe. The model introduces a parameter $c$ affecting the evolution equations, motivated by a modified Poisson algebra inspired by effective Loop Quantum Cosmology. Our analysis includes diverse background datasets such as Cosmic Chronometers, Pantheon+ Type Ia Supernovae (with and without the SH0ES calibration), SDSS, DESY6 and DESI Baryon Acoustic Oscillations, and background information of the Cosmic Microwave Background. We find that the model alleviates the Hubble tension in most of the dataset combinations, with cases reducing discrepancies to below $1\sigma$ when including SH0ES. However, the model exhibits minimal improvement in the overall fit when compared to $\Lambda$CDM, and Bayesian evidence generally favors the standard model. Theoretical foundations support this approach as a subtle adjustment to low-redshift dynamics, suggesting potential for further exploration into extensions of $\Lambda$CDM. Despite challenges in data fitting, our findings underscore the promise of small-scale modifications in reconciling cosmological tensions.
\end{abstract}


\maketitle

\section{Introduction}

Modern cosmology is affected by a number of puzzling questions, mainly concerning its dark
matter (DM)~\cite{Arbey:2021gdg} and dark energy (DE)~\cite{Frieman:2008sn} components. Furthermore, in recent years the data coming from different sources, like distance ladder measurements~\cite{Riess:2021jrx,Scolnic:2021amr,Scolnic:2023mrv,Anderson:2023aga,Jones:2022mvo,Anand:2021sum,Freedman:2021ahq,Uddin:2023iob,Huang:2023frr,Li:2024yoe,Pesce:2020xfe,Kourkchi:2020iyz,Schombert:2020pxm,Blakeslee:2021rqi,deJaeger:2022lit}, Cosmic Microwave Background Radiation (CMB)~\cite{Planck:2018vyg,ACT:2020gnv,SPT-3G:2022hvq}, Baryon Acoustic Oscillations (BAO)~\cite{eBOSS:2020yzd, adame2024desi}, and Weak Lensing and Galaxy Clustering~\cite{DES:2021wwk,DES:2018ufa,Harnois-Deraps:2024ucb,KiDS:2020ghu,Li:2023azi,Dalal:2023olq,Miyatake:2023njf,Sugiyama:2023fzm} have outlined the emergence of tensions in the determination of key cosmological parameters~\cite{Abdalla:2022yfr,Perivolaropoulos:2021jda,DiValentino:2020zio,DiValentino:2020vvd,DiValentino:2020srs}. The most statistically significant of such discrepancies is the so-called ``Hubble tension''~\cite{Verde:2019ivm}, and it is referring to the more than $5\sigma$ incompatibility in the value of the Hubble constant $H_0$ as determined by the SH0ES team and the Planck collaboration~\cite{Riess:2021jrx,Murakami:2023xuy,Breuval:2024lsv}. 

Various theoretical proposals to interpret this tension have been put forward in recent years, see for example Refs.~\cite{Vagnozzi:2018jhn,Poulin:2018cxd,DiValentino:2019ffd,Vagnozzi:2019ezj,DiValentino:2020vnx,Krishnan:2020obg,Jedamzik:2020krr,Colgain:2021pmf,Akarsu:2021fol,Naidoo:2022rda,Poulin:2023lkg,Adil:2023exv,Montani:2023ywn,montani2024kinetic,Giare:2023xoc,Vagnozzi:2023nrq,Meiers:2023gft,SolaPeracaula:2023swx,Pan:2023frx,Garny:2024ums,Giare:2024akf,Giare:2024smz,Anchordoqui:2024gfa,Akarsu:2024eoo,Lynch:2024hzh,Aboubrahim:2024spa,Pitrou:2023swx,Efstathiou:2023fbn,Gomez-Valent:2024tdb,Forconi:2023hsj,Silva:2024ift,Benisty:2024lmj,Montani:2024pou,Giare:2024syw,Ruchika:2024ymt,Mukherjee:2024akt} or the references in the review papers~\cite{DiValentino:2021izs,Kamionkowski:2022pkx,Khalife:2023qbu}. These scenarios have been mainly classified into two major categories~\cite{Knox:2019rjx}: early time solutions (which modify the expansion history before recombination) and late time solutions (which instead alter it after the recombination epoch). Unfortunately, none of them can resolve the Hubble tension, as investigated in Refs.~\cite{Jedamzik:2020zmd,Efstathiou:2021ocp,DiValentino:2022fjm,Vagnozzi:2023nrq}.
Another possibility is to adopt a modified theory of gravity, see for example Refs.~\cite{Odintsov:2020qzd, Schiavone:2022wvq, Montani:2023xpd}, where it is shown that metric $f(R)$-gravity theories could properly scale the Hubble function (as an effect of the non-minimal coupling that the scalar mode manifests with standard gravity in the so-called ``Jordan frame''). 
In particular, the analysis in~\cite{Schiavone:2022wvq} completes the theoretical arguments proposed in~\cite{Dainotti:2021pqg,Dainotti:2022bzg}, providing a 
representation of the observed variation of $H_0$ across a binned representation of the Pantheon sample for the Type Ia Supernovae (SNIa)~\cite{Pan-STARRS1:2017jku} (for similar studies see also~\cite{Krishnan:2020vaf,Kazantzidis:2020tko,Liu:2024vlt}). Despite not being statistically significant and facing criticism developed in~\cite{Brout:2020bbg}, the two studies~\cite{Dainotti:2021pqg, Dainotti:2022bzg} have highlighted an interesting and promising approach to addressing the Hubble tension problem.
In fact, as we also clarify below in this paper, if the explanation of the Hubble tension comes from a $z$-dependent rescaling of the $\Lambda$CDM-model dynamics, say we have to deal with 
an effective Hubble constant $\mathcal{H}_0(z)$, then the possibility to detect this effect could, in principle, concern also the analysis of nearby sources. With the expected increasing detection of new distant sources by the James Webb Space Telescope, 
we should regard this observational task 
as a challenging and intriguing perspective, to be extended also to Active Galactic Nuclei, 
QUASARS and Gamma Ray Bursts (for a discussion of the possible role of these sources as cosmological probes see~\cite{Dainotti:2022rea,Dainotti:2024aha}).

Here, we investigate a peculiar type of modified gravity, i.e., that one coming from the implementation of modified Poisson brackets to gravitational degrees of freedom, especially when the Minisuperspace of a cosmological model is concerned~\cite{Battisti:2009at, Battisti:2009zz, Barca:2021epy}. 
The idea is that, when we introduce cut-off physics effects in the early Universe quantum dynamics~\cite{Barca:2021qdn,Montani:2021rpe}, a theory of WKB modification of the standard dynamics is also observable when the quantum era of the Universe has ended. 
From a technical point of view, we can consider modified Heisenberg algebras, originating from fundamental theories, like String Cosmology~\cite{McAllister:2007bg}, Brane Cosmology~\cite{Papantonopoulos2002}, or Loop Quantum Cosmology~\cite{Ashtekar:2008zu}. These theories feature deformation parameters which can persist when the limit of a small Planck constant $\hbar$ is taken (due to their independence from $\hbar$ or their weak dependence as well). 
In this context, we discuss below an interesting candidate of modified Poisson algebra in relation to the solution of the Hubble tension, which leads to a revised cosmological dynamics containing three free parameters and possibly able to explain why the Cepheids calibrated Type Ia Supernovae (SN) data provide a different value of the Hubble constant compared to the Planck CMB value (the latter being naturally recovered as the redshift increases). 
We perform a data analysis of this model considering only the background parameters and low redshift data and we find that we can mildly alleviate the Hubble tension, only if a SH0ES-like prior on the Hubble constant is included.

The paper is structured as follows: in~\autoref{paradigm} we introduce the physical paradigm of the model, relying on suitable modified Poisson algebra, in~\autoref{dynamics} we derive the cosmological dynamics of the proposed model, providing the corresponding Hubble parameter for this scenario, in~\autoref{model} we investigate the appropriated parameter range to alleviate the Hubble tension, in~\autoref{data} we describe the data and methodology used for the analysis, in~\autoref{results} we present our findings, and in~\autoref{conclusions} we derive our conclusions.

\section{Physical paradigm}\label{paradigm}

In this paper, we consider a Hamiltonian theory, describing a one-dimensional problem in terms of a generalized coordinate $q$ and its conjugate momentum $p$, associated with an extended Poisson bracket of the form:

\begin{equation}
	\left\{ q\, ,\, p\right\} = 
	\sqrt{1 + f(\beta p^2)}
	\, , 
	\label{dei1}
\end{equation}
where $f$ is a generic function, while 
$\beta$ is a free parameter of the considered model with units so that $\beta p^2$ is a dimensionless quantity.

In order to reproduce the standard symplectic algebra in the limit of sufficiently small values of the momentum $p$, we have to require the following condition:

\begin{equation}
	\lim_{p\rightarrow 0} f(\beta p^2) = 0
	\, .
	\label{dei2}
\end{equation}

This kind of modified algebras is commonly implemented at a quantum level~\cite{Kempf:1993bq}, and their classical formulation makes sense in the spirit of the ``correspondence principle'': the classically modified Poisson brackets describe well the behavior of mean values for a localized wavepacket. When the generalized coordinate is identified, as shown below, with a gravitational degree of freedom~\cite{Barca:2021epy}, these theories can be thought of as modified Einsteinian dynamics, according to the idea that cut-off physics enter when high values of the momentum are considered.

One of the most studied formulations of the type described in Eq.~(\ref{dei1}) is referred to as the Generalized Uncertainty Principle, which corresponds to using the first term of a Taylor expansion in both the argument of $f$ and the square root~\cite{Kempf:1993bq, Kempf:1994su, Kempf:1994fv}. 
The interest in this specific case is due to its justification as a low-energy string dynamics~\cite{Amati:1987wq, Amati:1988tn} and a possible extension (which satisfies the Jacobi identities) can be found in~\cite{Maggiore:1993kv, Maggiore:1993rv, Segreto:2022clx}.

If we denote the system's Hamiltonian by $\mathcal{H}=\mathcal{H}(q, p)$, then the dynamics are associated with the Hamilton equations:

\begin{equation}
	\dot{q} = \frac{\partial \mathcal{H}}{\partial p}\sqrt{ 1 + f(\beta p^2)} \quad , \quad 
	\dot{p} = -\frac{\partial \mathcal{H}}{\partial q}\sqrt{1 + f(\beta p^2)}
	\, ,
	\label{dei3}
\end{equation}
where the dot denotes time differentiation. 
In what follows, we will consider a specific algebra corresponding to the function:

\begin{equation}
	f(\beta p^2) = 
	\frac{\beta p^2}{1 - \beta p^2}
	\, ,
	\label{dei4}
\end{equation}
which leads to the implementation of the following Poisson brackets:

\begin{equation}
	\left\{ q\, , p\right\} = 
	\sqrt{\frac{1}{1 - \beta p^2}}, \quad 
	-\frac{1}{\sqrt{\beta}} < 
	p < \frac{1}{\sqrt{\beta}}
	\,  
	\, .
	\label{dei5}
\end{equation}
This model can then be used in the phase space of a homogeneous and isotropic Universe to propose a cosmological model for it, as will be shown in the next section.

\section{Cosmological dynamics}\label{dynamics}

Let us consider a flat Robertson-Walker geometry~\cite{Montani:2009hju}, described by the line element (setting $c=1$):

\begin{equation}
	ds^2 = N^2(t)dt^2 - a^2(t) \delta_{ij}dx^idx^j
	\, , 
	\label{dei6}
\end{equation}
where $i,j=1,2,3$, $N$ is the lapse function (regulating the label time), and $a$ denotes the cosmic scale factor, which governs the Universe's expansion.

For this model, the Einstein-Hilbert action, in the presence of a matter energy density $\rho(a)$, takes the following expression:

\begin{equation}
	S = \int dt \left\{ - \frac{3}{\chi} \frac{a}{N} \dot{a}^2 + N \rho(a) a^3 \right\}
	\, , 
	\label{dei7}
\end{equation}
where $\chi$ is the Einstein constant and we set the fiducial volume on which the spatial integral is taken equal to unity. Hence, the momentum $p_a$, conjugate to the scale factor, reads:

\begin{equation}
	p_a = - \frac{6}{\chi} \frac{a}{N} \dot{a}
	\, , 
	\label{dei8}
\end{equation}
while the conjugate momentum to the lapse function is identically zero, according to time diffeomorphism invariance. It is in the phase space $\{ a\, ,\, p_a\}$ where we apply the modified Poisson algebra (\ref{dei5}), aiming to account in the classical Universe expansion for a reminiscence of the cut-off physics characterizing its primordial phase~\cite{Battisti:2008am, Battisti:2009at, Battisti:2009zz}.

Performing a Legendre transformation, we restate action (\ref{dei7}) in the Hamiltonian formulation as follows:

\begin{equation}
	S = \int dt \left\{ 
	p_a \dot{a} - N \left( - \frac{\chi}{12} \frac{p_a^2}{a} + \rho(a) a^3 \right) \right\}
	\, .
	\label{dei9}
\end{equation}

To this action, we add the definition of the cosmological fluid pressure as:

\begin{equation}
	p \equiv -\frac{1}{3a^2} \frac{d(\rho a^3)}{da} 
	\, , 
	\label{dei10}
\end{equation}
which effectively imposes energy-momentum tensor conservation on the adiabatic fluid.
Varying action (\ref{dei9}) with respect to the lapse function $N$, we obtain the Hamiltonian constraint:

\begin{equation}
	\frac{\chi}{12} p_a^2 = \rho a^4
	\, .
	\label{dei11}
\end{equation}
According to the first of the Hamilton equations (\ref{dei3}) and fixing the synchronous time by choosing $N=1$, we can write:

\begin{equation}
	\dot{a} = - \frac{\chi}{6} \frac{p_a}{a} \sqrt{\frac{1}{1 - \beta p_a^2}}
	\, .
	\label{dei12}
\end{equation}
It is worth stressing that this equation does not coincide with Eq. (\ref{dei8}) since we implemented the modified symplectic algebra in the Hamiltonian formulation (which is not applicable to the Lagrangian formalism).

We define the Hubble parameter as $H\equiv \dot{a}/a$. From Eq. (\ref{dei12}) and expressing $p_a$ via Eq. (\ref{dei11}), we arrive at the following modified Friedmann equation:

\begin{equation}
	H^2 = \frac{\chi}{3} \rho \frac{1}{1 - \frac{12\beta}{\chi} \rho a^4}
	\, .
	\label{dei13}
\end{equation}
In the next section, we will use the above expression to provide a physical interpretation of the Hubble tension.

We conclude this section by observing that, in the limit of small $\beta p_a^2$ (which is applicable to the late Universe), we have $\sqrt{1/(1 - \beta p_a^2)} \sim \sqrt{1 + \beta p_a^2} \sim 1 + \beta p_a^2/2$. Thus, it can be easily realized that similar results could be obtained both in the Generalized Uncertainty Principle (studied in~\cite{Kempf:1993bq}) and in its generalization of the form proposed in~\cite{Maggiore:1993kv, Maggiore:1993rv, Segreto:2022clx}.
The reason for choosing the algebra proposed in Eq. (\ref{dei5}) lies in its minimal modification to the primordial dynamics of the Universe. Despite its omission of the early-time evolution of the Universe, a brief discussion on the nature of the singularity is provided in Appendix \ref{appendix_a}.

\section{Versus the Hubble tension}\label{model}

In order to interpret our dynamical model, relying on Eq. (\ref{dei13}), we observe that the Universe energy density $\rho(z)$ can be expressed as follows (here $z \equiv \frac{1}{a} - 1$ is the redshift coordinate and we set the present-day value of the scale factor to unity):

\begin{equation}
	\rho(z) = \rho_0^{\text{crit}} \left[ \Omega_0^m (1 + z)^3 + \Omega_0^r (1 + z)^4 + \Omega_{\Lambda} \right]
	\, , 
	\label{dei14}
\end{equation}
where $\rho_0^{\text{crit}}$ denotes the Universe critical density at present-day, $\Omega_0^m$, $\Omega_0^r$, and $\Omega_{\Lambda}$ are the corresponding present-day values for the density parameters associated with (dark and baryonic) matter, radiation, and vacuum energy, respectively.
Hence, recalling that the following relation holds:

\begin{equation}
    H_*^2 = \frac{\chi}{3} \rho_0^{\text{crit}}
    \, , 
    \label{dei15}
\end{equation}
where $H_*$ is a fiducial value of the Hubble constant, we can restate Eq. (\ref{dei13}) in the form:

\begin{equation}
    H^2(z) = \frac{H_*^2}{f(z)} \left( \Omega_0^m (1 + z)^3 + \Omega_0^r (1 + z)^4 + \Omega_{\Lambda} \right)
    \, . 
    \label{dei16}
\end{equation}
The function $f(z)$ takes the explicit expression:

\begin{equation}
    f(z) \equiv 1 - c \left( \frac{\Omega_0^m}{1 + z} + \Omega_0^r + \frac{\Omega_{\Lambda}}{(1 + z)^4} \right)
    \, , 
    \label{dei17}
\end{equation}
where:

\begin{equation}
    c \equiv \frac{12\beta \rho_0^{\text{crit}}}{\chi}
    \, .
    \label{dei18}
\end{equation}
Since the sum of the density parameters is normalized to unity, we clearly have $H_0 \equiv H(z=0) = H_* / f(z=0)$. We stress that setting $a_0 \neq 1$ 
would simply correspond to rescaling 
$\beta$ as $\beta \rightarrow \beta a_0^4$.

\begin{figure}[h]
    \centering
    \includegraphics[width=0.52\textwidth]{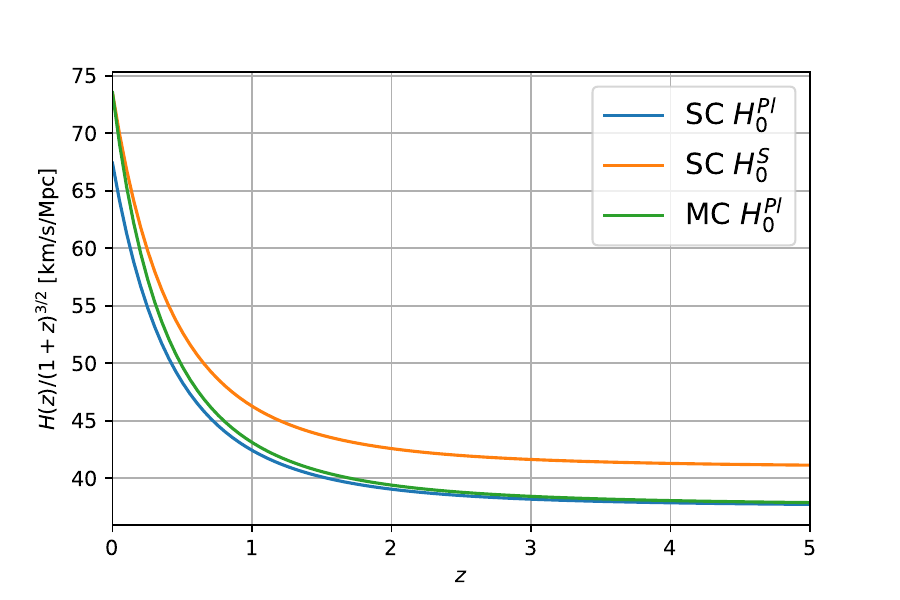}
    \caption{Yellow curve: standard $\Lambda$CDM cosmology (SC) with $H(z=0)=H_0^{S}$; Blue curve: SC with $H(z=0) = H_0^{Pl}$; Green curve: modified cosmology (MC) with $H_*=H_0^{Pl}$ and $c \sim 0.16$. In the three cases $\Omega_0^r$ has been neglected and $\Omega_0^m=0.31$ is assumed. The MC approaches the SC with $H(z=0)=H_0^{S}$ for small $z$ and the SC with $H(z=0) = H_0^{Pl}$ for $z$ larger than 1.}
    \label{fig:Hz}
\end{figure}

It is clear that, for $z \rightarrow \infty$, the function $f$ approaches the value $1 - c \Omega_0^r$, and this corresponds to a rescaling of the Einstein constant 
of the form $\chi_{\infty} \equiv 
\chi / (1 - c \Omega_0^r)$. For a $c$ that is at most of 
order unity (see below), this redefinition is not more than one part in $10^{-4}$, i.e. 
the order of magnitude of $\Omega_0^r$.
In order to study the impact of the presence of the function $f(z)$ in $H(z)$ versus the Hubble tension, 
we stress that:

\begin{equation}
    f(z=0) = 1 - c \quad , \, f(z \simeq 1100) \simeq 1
    \, ,
    \label{dei19}
\end{equation}
where we considered $c < 1$.
Thus, if we assume that the real value of $H_*$ is that measured by the Planck satellite (since it would correspond to $f=1$), i.e., 
$H_0^{Pl} \simeq 67.4 \, \mathrm{km} \, \mathrm{s}^{-1}\, \mathrm{Mpc}^{-1}$, 
then the Hubble constant value measured by SH0ES via SNIa has to be:

\begin{equation}
    H_0^{S} = \frac{H_0^{Pl}}{\sqrt{1 - c}} \simeq 73.5 \, \mathrm{km} \, \mathrm{s}^{-1}\, \mathrm{Mpc}^{-1}
    \, \rightarrow \, c \simeq 0.16
    \, .
    \label{dei20}
\end{equation}
In other words, we are stating that, phenomenologically, 
an effective $\mathcal{H}_0(z)$ has to be defined as 
$\mathcal{H}_0(z) \equiv \frac{H_0^{Pl}}{\sqrt{f(z)}}$. Figure \ref{fig:Hz} compares the modified cosmology of Eq. (\ref{dei16}) (with $c=0.16$ and $H_*=H_0^{Pl}$) with the standard cosmology for both $H(z=0) = H_0^{Pl}$ and $H(z=0) = H_0^{S}$. In the three cases, $\Omega_0^r$ has been neglected, and $\Omega_0^m=0.31$ is assumed.

The apparent variation of the Hubble constant should, in principle, be detectable at low-redshift sources, such as the SNIa distribution. This general perspective found inspiration from the binned analysis of the Pantheon sample~\cite{Pan-STARRS1:2017jku}, as developed in~\cite{Dainotti:2021pqg, Dainotti:2022bzg}. Although these results are not completely assessed and not identical to the predictions of our model, they encourage further efforts in this direction for data analysis.

\section{Data analysis methodology}\label{data}

In order to perform a proper quantitative data analysis, we neglect the quantity $\Omega_0^r$ in the functional form of $f(z)$ and set $\Omega_{\Lambda} = 1 - \Omega_0^m$. Furthermore, following the discussion presented in~\cite{Efstathiou:2021ocp}, we use the fitting function $E(z) \equiv H(z)/\sqrt{(1 + z)^3}$, i.e.:

\begin{equation}
    E(z) = \frac{H_* \sqrt{\Omega_0^m(1 + z)^3 + 1 - \Omega_0^m}}{\sqrt{(1 + z)^3 - c\left( \Omega_0^m(1 + z)^2 + \frac{1 - \Omega_0^m}{1 + z}\right)}}
    \, ,
    \label{dei21}
\end{equation}
This function contains three free parameters $H_*$, $\Omega_0^m$, and $c$, which are available for the fitting process. A case where curvature will be non-negligible will also be studied. In this case, the addition of the terms $\Omega^{k}_{0}(1+z)^2$ and $\Omega^{k}_{0}/(1+z)^2$ to Eq.~(\ref{dei16}) and Eq.~(\ref{dei17}), respectively, is paramount. We will draw upon Bayesian Statistics, particularly on the algorithm known as Nested Sampling~\cite{skilling2004nested}, to perform a parameter inference procedure. To this end, we make use of a modified version of the Bayesian Inference sampler named SimpleMC~\cite{simplemc}, a code used for calculations of distances and expansion rates with a given Friedmann equation, which implements the python library \texttt{dynesty}~\cite{Speagle:2019ivv} to perform parameter inference with the Nested Sampling algorithm.

To perform this procedure, and taking into account that we assume that our model only has notable repercussions in late-time cosmology, the following datasets will be used in different combinations:

\begin{itemize}
 \item Hubble parameter measurements of 31 Cosmic Chronometers~\cite{Jimenez:2003iv, Simon:2004tf, Stern:2009ep, Moresco:2012by, Zhang:2012mp, Moresco:2015cya, Moresco:2016mzx, Ratsimbazafy:2017vga}. This dataset will be referred to as ``CC''.
    
\item The Pantheon+ data release~\cite{Scolnic:2021amr} which consists of 1701 light curves of 1550 Type Ia Supernovae. When used it will be written as ``SN'' in the datasets and it spans a redshift of $0.01<z<2.26$.
    
\item Measurements of Baryon Acoustic Oscillations (BAO), which include the SDSS Galaxy Consensus, quasars, and Lyman-$\alpha$ forests~\cite{eBOSS:2020yzd}. The sound horizon is calibrated using BBN~\cite{Cooke:2013cba}. These datasets are comprehensively detailed in Table 3 of Ref.~\cite{eBOSS:2020yzd}. In this work, we will collectively refer to this set of measurements as ``BAO''.

\item The angular diameter distance measurement with BAO at $z_{\mathrm{eff}}=0.86$ from the Dark Energy Survey (DES)~\cite{abbott2024dark}. It will be written as ``DESY6'' in the datasets.
    
\item A prior for the Hubble Parameter $H_0 = 73.04\pm 1.04 \, \mathrm{km} \, \mathrm{s}^{-1}\, \mathrm{Mpc}^{-1}$ from the Hubble Space Telescope and the SH0ES program~\cite{Riess:2021jrx}. When using this prior we will write ``SH0ES'' along with the datasets used. An important note to make is that, when using this prior with our model, it must be used not on the free parameter $H_*$ (as would be the case for the standard model) but on $H_0=H_*/\sqrt{1-c}$, effectively influencing both parameters $H_*$ and $c$.
    
\item Measurements of BAO distances from the first year of the Dark Energy Spectroscopic Instrument (DESI)~\cite{DESI:2024mwx}. We will refer to this dataset as ``DESI''. Due to the risk of double-counting these will not be used in tandem with their SDSS counterpart.
    
\item Planck information, where the Cosmic Microwave Background (CMB) is considered as a ``BAO data point'' at a redshift of $z\sim 1100$, it is measured by the angular scale of the sound horizon at that epoch. As detailed in~\cite{BOSS:2014hhw}, the background-level information of the CMB can be encapsulated by the three parameters: $w_b$ (physical baryon density parameter), $w_{m}$ (physical matter density parameter), and $D_A(\sim 1100)/r_d$. This dataset will be referred to as ``Planck''.
\end{itemize}

The priors on the parameters are: $\Omega^{m}_{0} = [0.1, 0.5]$ for the matter density today, $h = [0.4, 0.9]$ for the dimensionless Hubble Parameter ($h = H_*/100$), $c = [0.0, 1.0]$ for our model-exclusive new parameter, and $\Omega^{k}_{0} = [-0.1, 0.3]$ for the energy density associated with the curvature of the Universe today. Finally, given that the Nested Sampling algorithm is being used for the parameter inference, we can perform model comparison between our model and $\Lambda$CDM. This is achieved by calculating the difference in their $-2\ln \mathcal{L}_{\rm max}$ (with $\mathcal{L}_{\rm max}$ being the maximum likelihood after performing parameter inference) and the difference in the natural logarithm of their Bayes' factors, which would take the form $\ln B_{\Lambda \text{CDM},i}$. 
To make evaluations of our model against the standard one, we will make use of the empirical Jeffrey's scale, particularly the convention from~\cite{Trotta:2008qt}. This scale tells us that the evidence against our model is: inconclusive if $0<\ln B_{\Lambda \text{CDM},i}<1.0$; weak if $1.0<\ln B_{\Lambda \text{CDM},i}<2.5$; moderate if $2.5<\ln B_{\Lambda \text{CDM},i}<5.0$; and strong if $5.0<\ln B_{\Lambda \text{CDM},i}$. Ideally, we would hope for a $\ln B_{\Lambda \text{CDM},i}<0.0$, which would indicate evidence in favor of our model despite being more complex (having more free parameters) than $\Lambda$CDM.

\section{Results}\label{results}

\begin{table*}[th!]
\caption{Summary of the mean values and the standard deviations (in parenthesis) for the parameters $h$, $H_0$, $c$, and $\Omega^{m}_{0}$ for different combinations of datasets. The last two columns correspond to the natural logarithm of the Bayes Factor $B_{\Lambda \text{CDM},i}$ which, if positive, indicates a preference for $\Lambda$CDM, and the $-2\Delta\ln \mathcal{L_{\rm max}} \equiv -2\ln( \mathcal{L_{\rm max}}_{,\Lambda \text{CDM}} / \mathcal{L_{\rm max,}}_i)$ which, if positive, indicates that $\Lambda$CDM has an improvement in the fit to the data.}
\footnotesize
\scalebox{0.89}{%
\begin{tabular}{cccccccc} 
\cline{1-8}\noalign{\smallskip}
\vspace{0.15cm}
Model & Datasets & $h$ & $H_0$ &  $\Omega^{m}_{0}$ &  $c$ &   $\ln B_{\Lambda \text{CDM},i}$  &  $-2\Delta\ln \mathcal{L_{\rm max}}$ \\
\hline
\vspace{0.15cm}
$\Lambda$CDM & CC+SN &  0.676 (0.028) & 67.58 (2.81) &   0.331 (0.017) & $0$ & $-$  &  $-$  \\
\vspace{0.15cm}
Our model & CC+SN &  0.644 (0.036) & 67.19 (5.77)  & 0.375 (0.041) & $<0.08$ &  2.2 (0.15)  & $-0.01$ \\
\hline
\vspace{0.15cm}
$\Lambda$CDM & CC+SN+SH0ES &  0.711 (0.018) &  71.14 (1.83)  &  0.322 (0.016) & $0$ & $-$  &  $-$  \\
\vspace{0.15cm}
Our model & CC+SN+SH0ES &  0.684 (0.029) &  70.95 (5.06)  & 0.361 (0.038) & $<0.09$ &  2.32 (0.17)  & $-0.02$ \\
\hline
\vspace{0.15cm}
$\Lambda$CDM & CC+SN+DESI+DESY6 &  0.675 (0.016) &  67.51 (1.65)  &  0.311 (0.012) & $0$ & $-$  &  $-$  \\
\vspace{0.15cm}
Our model & CC+SN+DESI+DESY6 &  0.669 (0.018) &  67.95 (2.71)  & 0.322 (0.016) & $<0.01$ &  3.41 (0.15)  & $-0.05$ \\
\hline
\vspace{0.15cm}
$\Lambda$CDM & CC+SN+DESI+DESY6+SH0ES &  0.696 (0.014) &  69.58 (1.39)  &  0.308 (0.012) & $0$ & $-$  &  $-$  \\
\vspace{0.15cm}
Our model & CC+SN+DESI+DESY6+SH0ES &  0.686 (0.016) &  69.87 (2.71)  & 0.322 (0.017) & $<0.04$ &  3.21 (0.16)  & $-0.21$ \\
\hline
 \vspace{0.15cm}
$\Lambda$CDM  & CC+SN+BAO+DESY6 &   0.688 (0.016) &  68.76 (1.62)  &   0.306 (0.012) & $0$ & $-$  &  $-$  \\
 \vspace{0.15cm}
Our model & CC+SN+BAO+DESY6 &  0.682 (0.018) &  69.22 (2.68)  & 0.319 (0.017) & $<0.03$ & 3.34 (0.19)  & $-0.03$ \\
\hline
\vspace{0.15cm}
$\Lambda$CDM & CC+SN+BAO+DESY6+SH0ES &   0.704 (0.014) &  70.37 (1.41)  &   0.305 (0.012) & $0$ & $-$  &  $-$  \\
\vspace{0.15cm}
Our model & CC+SN+BAO+DESY6+SH0ES &  0.696 (0.016) &  70.75 (2.57)  & 0.318 (0.017) & $<0.04$ &  3.22 (0.19)  & $-0.2$ \\
\hline
\vspace{0.15cm}
$\Lambda$CDM & CC+SN+BAO+DESY6+Planck &  0.677 (0.005) &  67.66 (0.54)  &  0.312 (0.007) & $0$ & $-$  &  $-$  \\
\vspace{0.15cm}
Our model   & CC+SN+BAO+DESY6+Planck &  0.678 (0.009) &  68.62 (1.67)  & 0.312 (0.011) & $<0.025$ &  3.09 (0.21)  & $-1.05$ \\
\hline
\vspace{0.15cm}
$\Lambda$CDM + $\Omega_{k}$=0.0105 (0.012) & CC+SN+BAO+DESY6 &   0.684 (0.017) &  68.77 (2.08)  &   0.303 (0.012) & $0$ & $-$  &  $-$  \\
\vspace{0.15cm}
Our model + $\Omega_{k}$=0.011 (0.011) & CC+SN+BAO+DESY6 &  0.678 (0.018) &  69.28 (3.13)  & 0.316 (0.017) & $<0.03$ &  3.31 (0.21)  & $-0.14$ \\
\hline
\vspace{0.15cm}
$\Lambda$CDM + $\Omega_{k}$=0.008 (0.011) & CC+SN+BAO+DESY6+SH0ES &   0.702 (0.014) &  70.45 (1.83)  &  0.302 (0.012) & $0$ & $-$  &  $-$  \\
\vspace{0.15cm}
Our model + $\Omega_{k}$=0.008 (0.012) & CC+SN+BAO+DESY6+SH0ES &  0.694 (0.017) &  70.88 (3.15)  & 0.316 (0.017) & $<0.03$ &  2.98 (0.19)  & $-0.11$ \\
\hline
\vspace{0.15cm}
$\Lambda$CDM + $\Omega_{k}$=0.003 (0.003) & CC+SN+BAO+DESY6+Planck &  0.687 (0.011) &  68.71 (1.08)  &   0.304 (0.011) & $0$ & $-$  &  $-$  \\
\vspace{0.15cm}
Our model + $\Omega_{k}$=0.0 (0.003) & CC+SN+BAO+DESY6+Planck &  0.679 (0.013) &  68.74 (2.12)  & 0.312 (0.012) & $<0.025$ &  3.39 (0.22)  & $-0.12$ \\
\hline
\hline
\end{tabular}}
\label{tabla_evidencias}
\end{table*}

The general results after performing parameter inference can be found in Table \ref{tabla_evidencias}. One important thing to note is that every case, with the exception of the one where only Cosmic Chronometers and Pantheon+ are used, present a slightly higher value of $H_0$ than the standard model (it is worth noting that we refer to $H_0$ as the value of $H(z=0)$, not the fiducial value of the Hubble constant $H_*$). This is naturally attributed to the extra freedom that the new parameter $c$ provides, but it comes at the cost of a larger uncertainty in its inference. This increase in the uncertainty is expected as, for our model, the value of $H_0$ is not a free parameter, like in $\Lambda$CDM's case, but a derived quantity whose value depends on every other inferred parameter of the model, effectively increasing its standard deviation. Regarding the case with CC+SN, the error bars on the $H_0$ are very relaxed, so the addition of the SH0ES prior brings $H_0=70.9\pm5.06 \, \mathrm{km} \, \mathrm{s}^{-1}\, \mathrm{Mpc}^{-1}$ at 68\% CL for CC+SN+SH0ES in our model.

\begin{figure*}[t!]
    \centering
    \makebox[11cm][c]{
    \includegraphics[trim = 0mm  0mm 0mm 0mm, clip, width=10.cm, height=6.cm]{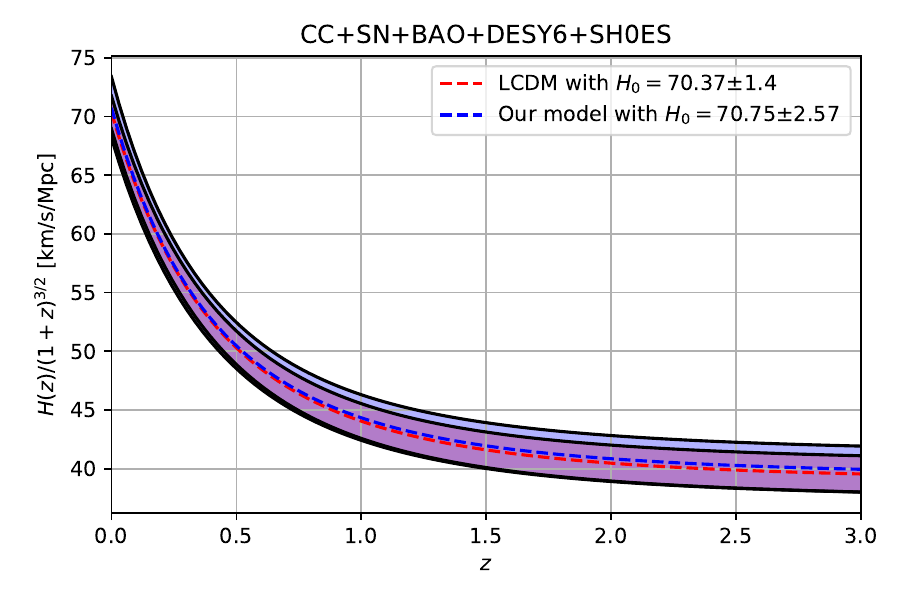}
    \includegraphics[trim = 0mm  0mm 0mm 0mm, clip, width=10.cm, height=6.cm]{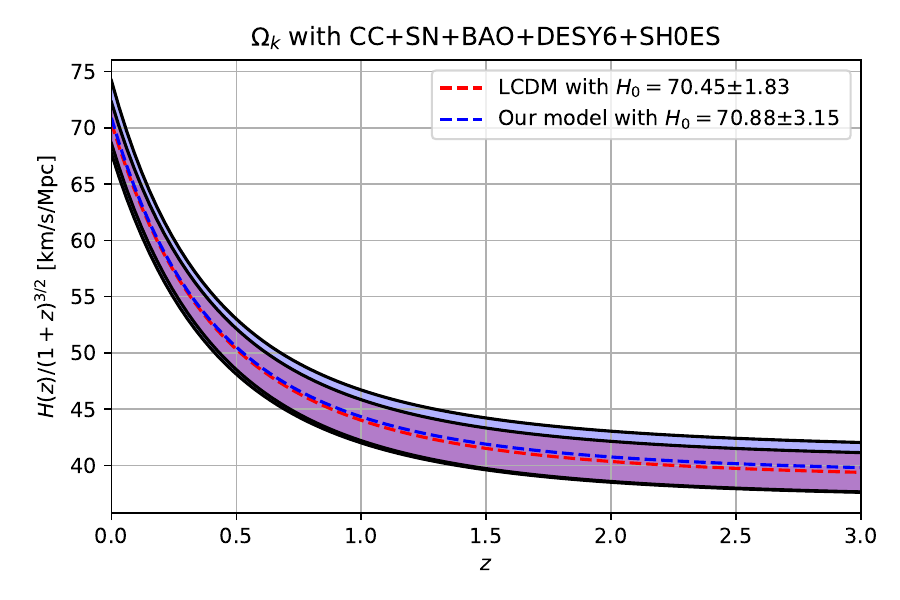}
    }
    \caption{Comparison of the $H(z)/(1+z)^{3/2}$ as a function of $z$, between $\Lambda$CDM and our model for the dataset combination CC+SN+BAO+DESY6+SH0ES for the case without curvature (left panel) and with curvature (right panel). Our model accommodates a slightly higher value of the Hubble constant in both cases.}
    \label{fig:hz_comparison_1sigma}
\end{figure*}

Two cases stand out from the bunch and they are plotted in Fig. \ref{fig:hz_comparison_1sigma}. These correspond to the cases where the dataset combination CC+SN+BAO+DESY6+SH0ES was used. One of them was allowed to have a parameter for a contribution to the total energy of the Universe in the form of spatial curvature. These two cases give a value of the Hubble constant that is $H_0=70.75\pm2.5 \, \mathrm{km} \, \mathrm{s}^{-1}\, \mathrm{Mpc}^{-1}$ ($H_0=70.88\pm3.1 \, \mathrm{km} \, \mathrm{s}^{-1}\, \mathrm{Mpc}^{-1}$) for the case without (with) $\Omega^{k}_{0}$ at 68\% CL. Therefore, they present an alleviation of the Hubble Tension with Planck assuming the standard $\Lambda$CDM model by reducing it to 1.31$\sigma$ (without $\Omega^{k}_{0}$) and 1.11$\sigma$ (with $\Omega^{k}_{0}$), and the tension with SH0ES to 0.84$\sigma$ (without $\Omega^{k}_{0}$) and 0.66$\sigma$ (with $\Omega^{k}_{0}$). In all the cases, $\Omega^{k}_{0}$ is slightly positive, but in agreement with a flat universe within 1$\sigma$.

Despite the achievements in reducing the Hubble Tension, we cannot ignore both the fit to the data and our Bayes' factor, represented in Table \ref{tabla_evidencias} by $-2\Delta\ln \mathcal{L_{\rm max}}$ and $\ln B_{\Lambda \text{CDM},i}$, respectively. 
The overall fit to the data for every case is better when compared against $\Lambda$CDM, but by a negligible amount. Only the case with CC+SN+BAO+DESY6+Planck reaches slightly above the 1$\sigma$ level, while every other case stays well below this margin. We can attribute this minor improvement to the low value inferred for our new parameter $c\sim 0.01-0.03$ (at least when compared to our theoretical expectation of $c\sim0.16$, which seems to be heavily disfavored by the data as explained in Appendix~\ref{appendix_cfixed}), resulting in a slightly altered version of $\Lambda$CDM at low-redshifts.
Regarding the Bayes' factor, we find that all cases present weak (at best) and moderate (at worst) evidence against them. This result is expected since, as the $-2\Delta\ln \mathcal{L_{\rm max}}$ shows, even with the addition of an extra parameter our model does not provide enough of an improvement when trying to fit/explain the datasets used. These two factors (the extra free parameter and the lack of a significant difference in $-2\ln \mathcal{L_{\rm max}}$ when compared with the standard model) are penalized by the Bayesian evidence.

\section{Concluding remarks}\label{conclusions}

In this paper, we analyzed a revised cosmological dynamics as a possible candidate to alleviate the Hubble tension. The theoretical ground of our reformulation of the late Universe dynamics consists of a modified Poisson algebra associated with the late Universe evolution. 
Actually, we clarified (see the Appendix) how the early Universe dynamics are weakly affected by the considered scenario, while in the DE-dominated Universe the Friedmann equation significantly deviates from the standard ($\Lambda$CDM) behavior. 
From a theoretical point of view, the justification for dealing with a modified Poisson algebra affecting the late-time cosmology comes from the idea of an infrared implication of a cut-off physics, 
in close analogy to effective Loop Quantum Cosmology \cite{Ashtekar:2008zu,Giovannetti:2021vqj}. In other words, we considered a limit in which the fundamental constant $\hbar$ goes to zero, while the cut-off parameter remains finite. Thus, we deal with a sort of modified gravity formulation for the cosmological dynamics, emerging as a WKB effect of a quantum deformation of the physics in the Planck era of the Universe \cite{CanQuanGravCianf}.

The concrete implication of our formulation is a $z$-dependent factor, scaling the usual $\Lambda$CDM model. 
It has been clearly argued how such a rescaling factor tends to unity as the redshift increases, so that the Universe evolution is reconciled to a 
Hubble parameter whose value is in line with the expectations from the Planck satellite measurements. On the contrary, as shown in figure (\ref{fig:Hz}), this factor, near $z \simeq 0$, enhances the value of 
$H(z)$ permitting to pass, in principle, from a Planck value of the Hubble constant to that one detected by the SH0ES collaboration. 
Our modified Friedmann equation contains only one additional free parameter with respect to a standard $\Lambda$CDM model, i.e. the constant $c$. In order for the tension to be fully addressed we determined that the value of this parameter has to be about $0.16$, when the parameter $H_*$ has the Planck value.

After establishing the theoretical background of our model, we proceeded to perform a reconstruction of its modified Friedmann equation, based on some of the more relevant background low-redshift data, i.e., Cosmic Chronometers, Type Ia Supernovae, and BAO distances. The results showed an alleviation of the Hubble tension, with some cases reducing this tension below $1\sigma$ when SH0ES is included, forcing $H_0$ to take higher values. Nevertheless, its fit to the data showed little to no improvement and the Bayesian evidence still slightly favors the standard model.

Undeterred by the somewhat negative results encountered in the data-analysis part of this work, it is encouraging that a small modification to the standard model at low redshifts presents great potential for alleviating (or even solving) the Hubble tension. We expect this result to motivate the study of similar types of extensions of the $\Lambda$CDM model.

\section*{Acknowledgments}

E.D.V. is supported by a Royal Society Dorothy Hodgkin Research Fellowship. We also want to thank the Unidad de C\'omputo of ICF-UNAM for their assistance in the maintenance and use of the computing equipment. 

\vspace{0,5cm}

\appendix

\section{Bouncing cosmology} \label{appendix_a}

\begin{figure}[h]
    \centering
    \includegraphics[width=0.5\textwidth]{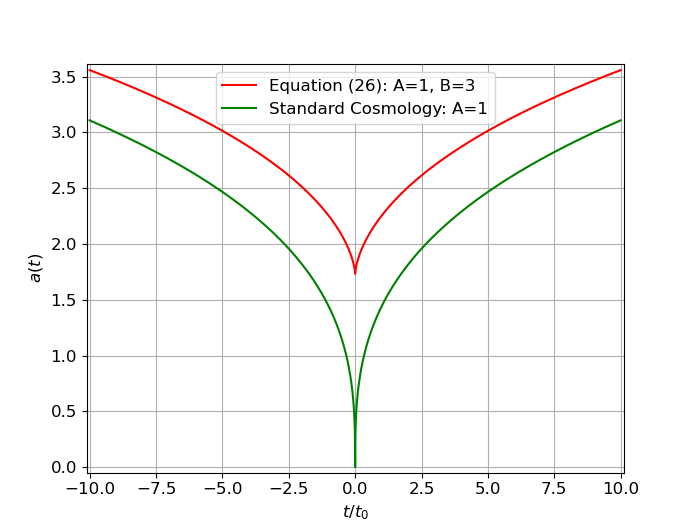}
    \caption{Red curve: solution of the modified Friedmann equation with $A=1$ and $B=3$. Green curve: standard cosmology solution with $A=1$.}
    \label{fig:Ap1}
\end{figure}

The Bianchi identity for the Einstein tensor implies Eq. (\ref{dei10}), which we restate here in the form:

\begin{equation}
    \dot{\rho} = -3H \left( \rho + p \right)
    \, .
    \label{a1}
\end{equation}

In correspondence to the equation of state $p = w\rho$, Eq. (\ref{a1}) provides the following expression for $\rho (a)$:

\begin{equation}
    \rho = \frac{\mu^2_w}{a^{3(1+w)}} 
    \, , 
    \label{a2}
\end{equation}

where $\mu^2_w$ denotes a positive quantity. Substituting the expression above into Eq. (\ref{dei13}) yields

\begin{equation}
    H^2 = \frac{A}{a^{3(1+w)}}\frac{1}{1 - \frac{B}{a^{3w-1}}}
    \, , 
    \label{a3}
\end{equation}

where we set $A \equiv \chi \mu^2_w / 3$ 
and $B \equiv 12\beta \mu_w^2 / \chi$.

Recalling that $H = \dot{a}/a$, Eq. (\ref{a3}) provides the solution $a(t)$, which governs the Universe evolution, in both the collapsing and expanding branch, corresponding to the negative and positive root of $H$, respectively.

As far as the equation of state parameter $w$ remains in the interval $0 \le w < 1/3$, no change emerges in the approach of the Universe toward the initial singularity. Also in the case of a radiation contribution, having $w = 1/3$, the singularity is reached like in the standard cosmology, apart from a weak rescaling of the Einstein constant, as previously discussed.

Instead, the case of matter with $1/3 < w \le 1$ seems to possess a new feature, i.e., the emergence of a minimal value of the scale factor for $a = a_{\min} \equiv B^{1/(3w - 1)}$. Now, in order to elucidate that such a configuration exactly coincides with the initial singularity, we analyze the relevant case $w = 1$, mimicking the contribution of the kinetic term of the inflaton field.

For $w = 1$ the positive and negative roots of Eq. (\ref{a3}) read as:

\begin{equation}
    \dot{a} = \pm \frac{\sqrt{A}}{a}\frac{1}{\sqrt{a^2 - B}}
    \, , 
    \label{a4}
\end{equation}

whose solutions stand as follows:

\begin{equation}
    a(t) = \pm \sqrt{B + (3\sqrt{A} t/t_0)^{2/3}}
    \, . 
    \label{a5}
\end{equation}

Comparing these solutions with the corresponding ones for standard cosmology, i.e., $a(t) = \pm (3\sqrt{A} t/t_0)^{1/3}$ (see Fig. \ref{fig:Ap1}), we see that they are obtained from Eq. (\ref{a5}) when $B = a_{\min}^2 \rightarrow 0$. Thus, we can conclude that also for $1/3 < w \le 1$, the solution of the modified Friedmann equation (\ref{a3}) does not introduce new physics in the early Universe.

\section{The theoretical value $c=0.16$}\label{appendix_cfixed}

\begin{table*}[th!]
\caption{Summary of the mean values and the standard deviations (in parenthesis) for the parameters $h$, $H_0$, $c$, and $\Omega^{m}_{0}$ for different combinations of datasets. The last two columns correspond to the natural logarithm of the Bayes Factor $B_{\Lambda \text{CDM},i}$ which, if positive, indicates a preference for $\Lambda$CDM, and the $-2\Delta\ln \mathcal{L_{\rm max}} \equiv -2\ln( \mathcal{L_{\rm max}}_{,\Lambda \text{CDM}} / \mathcal{L_{\rm max,}}_i)$ which, if positive, indicates that $\Lambda$CDM has an improvement in the fit to the data.}
\footnotesize
\scalebox{0.89}{%
\begin{tabular}{cccccccc} 
\cline{1-8}\noalign{\smallskip}
\vspace{0.15cm}
Model & Datasets & $h$ & $H_0$ &  $\Omega^{m}_{0}$ &  $c$ &   $\ln B_{\Lambda \text{CDM},i}$  &  $-2\Delta\ln \mathcal{L_{\rm max}}$ \\
\hline
\vspace{0.15cm}
$\Lambda$CDM & CC+SN &  0.676 (0.028) & 67.58 (2.81) &   0.331 (0.017) & $0$ & $-$  &  $-$  \\
\vspace{0.15cm}
Our model ($c$ fixed) & CC+SN &  0.613 (0.025) &  66.84 (2.74) & 0.422 (0.019) & $0.16$ &  1.49 (0.21)  & $2.51$ \\
\hline
\vspace{0.15cm}
$\Lambda$CDM & CC+SN+SH0ES &  0.711 (0.018) &  71.14 (1.83)  &  0.322 (0.016) & $0$ & $-$  &  $-$  \\
\vspace{0.15cm}
Our model ($c$ fixed) & CC+SN+SH0ES &  0.679 (0.018) &  74.04 (1.91) & 0.406 (0.017) & $0.16$ &  6.52 (0.21)  & $6.32$ \\
\hline
\vspace{0.15cm}
$\Lambda$CDM & CC+SN+DESI+DESY6 &  0.675 (0.016) &  67.51 (1.65)  &  0.311 (0.012) & $0$ & $-$  &  $-$  \\
\vspace{0.15cm}
Our model ($c$ fixed) & CC+SN+DESI+DESY6 &  0.632 (0.015) &  68.94 (1.64) & 0.381 (0.013) & $0.16$ &  6.05 (0.24)  & $6.21$ \\
\hline
\vspace{0.15cm}
$\Lambda$CDM & CC+SN+DESI+DESY6+SH0ES &  0.696 (0.014) &  69.58 (1.39)  &  0.308 (0.012) & $0$ & $-$  &  $-$  \\
\vspace{0.15cm}
Our model ($c$ fixed) & CC+SN+DESI+DESY6+SH0ES &  0.664 (0.014) &  72.43 (1.51)  & 0.374 (0.013) & $0.16$ &  10.71 (0.24)  & $10.5$ \\
\hline
 \vspace{0.15cm}
$\Lambda$CDM  & CC+SN+BAO+DESY6 &   0.688 (0.016) &  68.76 (1.62)  &   0.306 (0.012) & $0$ & $-$  &  $-$  \\
 \vspace{0.15cm}
Our model ($c$ fixed) & CC+SN+BAO+DESY6 &  0.682 (0.018) &  70.38 (1.67)  & 0.379 (0.014) & $0.16$ & 6.45 (0.25)  & $6.29$ \\
\hline
\vspace{0.15cm}
$\Lambda$CDM & CC+SN+BAO+DESY6+SH0ES &   0.704 (0.014) &  70.37 (1.41)  &   0.305 (0.012) & $0$ & $-$  &  $-$  \\
\vspace{0.15cm}
Our model ($c$ fixed) & CC+SN+BAO+DESY6+SH0ES &  0.673 (0.014) &  73.43 (1.49)  & 0.375 (0.014) & $0.16$ &  9.98 (0.24)  & $10.11$ \\
\hline
\vspace{0.15cm}
$\Lambda$CDM & CC+SN+BAO+DESY6+Planck &  0.677 (0.005) &  67.66 (0.54)  &  0.312 (0.007) & $0$ & $-$  &  $-$  \\
\vspace{0.15cm}
Our model  ($c$ fixed) & CC+SN+BAO+DESY6+Planck &  0.632 (0.006) &  69.01 (0.62) & 0.361 (0.009) & $0.16$ &  7.22 (0.24)  & $8.02$ \\
\hline
\vspace{0.15cm}
$\Lambda$CDM + $\Omega_{k}$=0.0105 (0.012) & CC+SN+BAO+DESY6 &   0.684 (0.017) &  68.77 (2.08)  &   0.303 (0.012) & $0$ & $-$  &  $-$  \\
\vspace{0.15cm}
Our model ($c$ fixed) + $\Omega_{k}$=0.014 (0.011) & CC+SN+BAO+DESY6 &  0.641 (0.015) &  70.56 (2.11) & 0.375 (0.014) & $0.16$ &  5.79 (0.24)  & $5.2$ \\
\hline
\vspace{0.15cm}
$\Lambda$CDM + $\Omega_{k}$=0.008 (0.011) & CC+SN+BAO+DESY6+SH0ES &   0.702 (0.014) &  70.45 (1.83)  &  0.302 (0.012) & $0$ & $-$  &  $-$  \\
\vspace{0.15cm}
Our model ($c$ fixed) + $\Omega_{k}$=0.011 (0.011) & CC+SN+BAO+DESY6+SH0ES &  0.671 (0.013) &  73.65 (1.94)  & 0.371 (0.014) & $0.16$ &  9.54 (0.24)  & $10.12$ \\
\hline
\vspace{0.15cm}
$\Lambda$CDM + $\Omega_{k}$=0.003 (0.003) & CC+SN+BAO+DESY6+Planck &  0.687 (0.011) &  68.71 (1.08)  &   0.304 (0.011) & $0$ & $-$  &  $-$  \\
\vspace{0.15cm}
Our model ($c$ fixed) + $\Omega_{k}$=-0.001 (0.003) & CC+SN+BAO+DESY6+Planck &  0.629 (0.011) &  68.62 (1.22) & 0.364 (0.012) & $0.16$ &  7.83 (0.24)  & $8.21$ \\
\hline
\hline
\end{tabular}}
\label{tabla_evidencias2}
\end{table*}

To test the validity of our theoretical prediction $c=0.16$, we also performed some tests where this parameter was fixed to this value. The results of these runs can be found in Table~\ref{tabla_evidencias2}, and it is evident that they are strongly rejected by the datasets used. Every single $-2\Delta\ln \mathcal{L}_{\rm max}$ presents preference for $\Lambda$CDM, some cases even surpassing $3\sigma$ preference for it. Not to mention the Bayes' factor, which shows a moderate to strong evidence in favor of the standard model despite presenting the same number of free parameters as our model with $c$ fixed.

The problems do not stop there; when forcing $c$ to behave in such a way, we recover larger values of the matter density parameter $\Omega_{0}^{m}$ and lower values for $h$, which would increase the tension with the Planck value even more at large redshifts. These results highlight the importance of data in a model-selection procedure.

\bibliography{biblio}
\end{document}